\documentclass[12pt]{article}
\usepackage{epsf}
\def\dst{\displaystyle}
\def\f{\frac}
\def\b{\beta}

\def\m{\mu}

\def\t{\theta}
\def\p{\partial}
\def\e{\varepsilon}
\def\eps{\varepsilon}
\def\be{\begin{equation}}
\def\ee{\end{equation}}
\def\bea{\begin{eqnarray}}
\def\eea{\end{eqnarray}}

\def\ba{\begin{array}}
\def\ea{\end{array}}
\def\bea{\begin{eqnarray}}
\def\eea{\end{eqnarray}}

\def\l{\left}
\def\r{\right}
\def\Det{\mathrm{Det}}

\begin{document}
\begin{titlepage}
\begin{center}
{\Large \bf William I. Fine Theoretical Physics Institute \\
University of Minnesota \\}
\end{center}
\vspace{0.2in}
\begin{flushright}
FTPI-MINN-08/37 \\
UMN-TH-2719/08 \\
September 2008 \\
\end{flushright}
\vspace{0.3in}
\begin{center}
{\Large \bf Breaking of a metastable string at finite temperature.
\\}
\vspace{0.2in}
{\bf A. Monin \\}
School of Physics and Astronomy, University of Minnesota, \\ Minneapolis, MN
55455, USA, \\
and \\
{\bf M.B. Voloshin  \\ }
William I. Fine Theoretical Physics Institute, University of
Minnesota,\\ Minneapolis, MN 55455, USA \\
and \\
Institute of Theoretical and Experimental Physics, Moscow, 117218, Russia
\\[0.2in]
\end{center}

\begin{abstract}

We consider the phase transition of a string with tension $\eps_1$ to a string
with a smaller tension $\eps_2$ at finite temperature. For sufficiently small
temperatures the transition proceeds through thermally catalyzed quantum
tunneling, and we calculate in arbitrary number of dimensions the thermal
catalysis factor. At $\eps_2=0$ the found formula for the
decay rate also describes a break up of a metastable string into two pieces.

\end{abstract}

\end{titlepage}

\section{Introduction}

A string-like configuration with string tension $\e_1$ can be metastable with
respect to either a complete breaking\cite{Vilenkin,Preskill} or a transition to
a string of a lower tension $\e_2$, with $\e_2 < \e_1$~\cite{Shifman}. The
metastability arises through existence of an energy barrier due to the emergence
in the process of two interfaces (the ends of the original string) each having
mass $\mu$. At zero temperature the breaking of the string proceeds due to
quantum tunneling and nucleation of a critical gap in the original string with
the length $\ell_c=2 \mu/(\e_1-\e_2)$, at which length the energy gain due to
the lower tension in the gap compensates for the energy $2 \mu$ required for the
formation of the interfaces. Once nucleated, the critical gap expands and
converts the whole length of the original string into either a string with the
lower tension, or into `nothing', which formally corresponds to the limit $\e_2
\to 0$. This mechanism of the string transition bears strong
similarity\cite{Vilenkin,Preskill,Shifman} to the decay of false
vacuum\cite{vko,Coleman} and to the Schwinger process\cite{Schwinger} of
creation of pairs of charged particles by electric field. However there is an
important difference between the string transition and the latter processes.
Namely, the transverse waves on the string correspond to a presence of massless
degrees of freedom, which substantially affect the probability of nucleation of
the critical gap. The effect of the transverse degrees of freedom  has recently
been calculated\cite{mv08} in terms of the preexponential factor in the rate of
the nucleation at zero temperature of critical gaps per length of the original
string in a $d$-dimensional theory:
\be
{d \Gamma \over d \ell}={\eps_1-\eps_2 \over 2 \pi} \, \left [ F \left ( {\eps_2
\over \eps_1} \right ) \right ]^{d-2} \, \exp \left ( -
\, {\pi \, \mu_R^2 \over \eps_1-\eps_2} \right )~,
\label{gf}
\ee
with $\mu_R$ being the renormalized value of the mass of the interface that
includes the effects
of the transverse
motion of the adjacent part of the string, and $F$ is the factor contributed in
the rate by each of the $(d-2)$ transverse dimensions and given by
\be
F(\rho)=\sqrt{{1+\rho \over 2}} \, \Gamma \left ({2 \over 1-\rho} \right ) \,
\left ( {1+\rho \over 1- \rho} \right )^{1+\rho \over 1- \rho} \, \exp \left (
{1+\rho \over 1- \rho} \right ) \ \left ( 2 \pi \, {1+\rho \over 1- \rho} \right
)^{-1/2}~.
\label{frho}
\ee

In this paper we consider the thermal effect in the rate of the string
transition at low temperatures. This effect further exposes the difference
between the breaking of a string on one side and the false vacuum decay and the
Schwinger process on the other. Namely, as long as the temperature $T$ is lower
than the inverse of the critical length, $T < 1/\ell_c$ the thermal effects in
the latter processes is exponentially suppressed as $\exp(-m/T)$ with $m$ being
the lowest scale for particle masses in the theory, and these effects are
essentially due to the (exponentially small) presence of massive particles in
the thermal equilibrium\cite{Garriga94}. In the case of a string, however, the
transverse waves on the string are massless so that their excitation has no
suppression by the mass at arbitrarily low temperature. The thermal excitations
of these waves create fluctuations in the distribution of energy in the string
which catalyze the nucleation of the critical gap. Clearly, at $T \ll 1/\ell_c$
the typical wavelength of the thermal waves $\lambda \sim 1/T$ is large in
comparison with the critical length $\ell_c$, and the thermal effect in the rate
is quite small, although not exponentially small. We find, as a result of the
calculation to be presented in this paper, that the leading low temperature
correction in the nucleation rate is given by the thermal catalysis factor
\be
K=\left \{ 1 + {\pi^8 \over 450}\, \left ( {\e_1-\e_2 \over 3 \e_1-\e_2} \right
)^2 \, \l(\f{\ell_c T}{2}\r)^8 + O\left[(\ell_c T)^{12} \right ] \right
\}^{d-2}~.
\label{kf}
\ee
We also find that as $T$ approaches $1/\ell_c$ the catalysis factor develops a
singularity at $\ell_c T =1$. At still higher temperatures the considered here
string transition behaves, in a sense, similarly to the false vacuum
decay\cite{Garriga94}, namely the regime of the transition changes to a
different tunneling trajectory, so that the temperature dependence appears in
the semiclassical exponential factor in Eq.(\ref{gf}), rather than in the
preexponential term.

The material in this paper is organized as follows. In the next Section we
formulate the problem in terms of periodic configurations in Euclidean
space-time, and in Sect.~3 we calculate the relevant path integrals and find the
general formula for the catalysis factor in terms of expansion in powers of
$\ell_c T$. In Sect.~4 we further analyze the general formula and find the first
terms in the low temperature expansion of the catalysis factor, and also
calculate this factor numerically for the temperatures approaching the critical
point at $\ell_c T=1$. Finally, Sect.~5 contains a summary of our results.

\section{Euclidean-space calculation}
The tunneling trajectory describing the decay of a metastable state can be found
as a classical configuration, called ``the bounce", in the Euclidean
space-time\cite{Coleman}, and the exponential factor for the decay rate is given
by the classical action on the bounce. The fields in the bounce configuration
approach their values in the metastable state at the boundaries of the
space-time and they approach the final state values inside the bounce. 

In the discussed here problem of the string transition the bounce is found in
terms of the effective low-energy Nambu-Goto action for the states of the
string.  For two strings with tensions $\e_1$ and $\e_2$ and a particle with
mass $\m$ propagating along the interface of the strings one can write the
action in the form
\be
S=\mu \, P + \eps_1 \, A_1 + \eps_2 \, A_2~,
\label{a0}
\ee
with $P$ being the perimeter of the interface and $A_i$ being the corresponding
string's world surface area. The approximation of the action of the system with
the effective low-energy one is applicable as long as massive internal degrees
of freedom of the strings are not excited. This implies in particular that the
approximation is applicable only as long as the thickness of the strings can be
neglected comparing to other length scales in the problem, or equivalently only
as long as all the momenta in the problem are small compared to the inverse
thickness of the string. If the thickness of the string is $r_0$ and the mass
scale associated with it is $M_0=1/r_0$ then we can write the condition of
applicability of the approach presented in the form
\be
l\gg r_0,\,~~~~~k\ll M_0
\label{appl_cond}
\ee
We consider initially a string with tension $\e_1$ with very large length $X$
stretched along the $x$ direction. At zero temperature the stationary
configurations for the action (\ref{a0}) are the trivial one, which is a flat
world surface of the string lying in $(t,x)$ plane, and the bounce
configuration, which has a disk filled with the lower phase of the string with
the radius
\be
R=\f{\m}{\e_1-\e_2}~,
\ee
as shown in Fig.~1. One can readily notice that the length of the critical gap
is the diameter of the disk, $\ell_c=2 \, R$.
\begin{figure}[ht]
  \begin{center}
    \leavevmode
    \epsfxsize=12cm
    \epsfbox{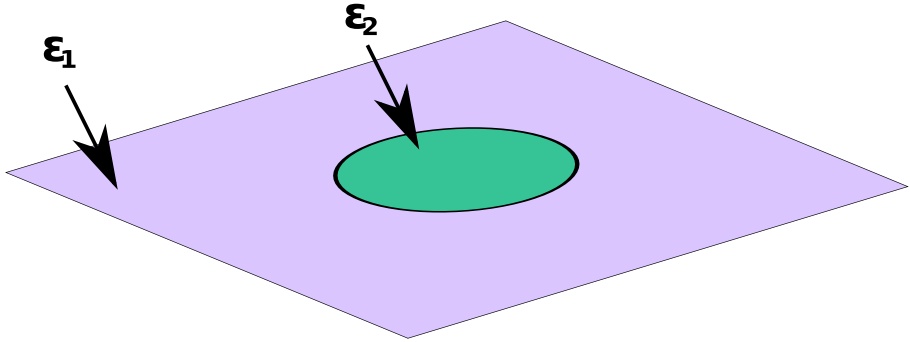}
    \caption{Bounce configuration for zero temperature}
  \end{center}
\label{bounce_T_zero}
\end{figure}
The condition (\ref{appl_cond}) for applicability of the effective action
(\ref{a0}) requires that
\be
\f{R}{r_0}\equiv M_0 \, R = {M_0 \, \mu \over  \eps_1-\eps_2} \gg 1~.
\label{rineq}
\ee

The rate of the critical gap nucleation is then found\cite{Stone,Coleman,Callan}
by calculating the path integral ${\cal Z}_{12}$ around the bounce and comparing
it with the path integral ${\cal Z}_1$ around the trivial  configuration:
\be
{d \Gamma \over d \ell}= {1 \over {\cal A}} \, \mathrm{Im}\f{{\cal
Z}_{12}}{{\cal Z}_{1  }}~,
\label{rz}
\ee
where ${\cal A}$ is the total area of the world surface for the string. It can
also be reminded that, as explained in great detail in Ref.\cite{Callan},
the imaginary part of ${\cal Z}_{12}$ arises from one negative mode at the
bounce configuration, and that due to two translational zero modes the
numerator in Eq.(\ref{rz}) is proportional to the total space time area ${\cal
A}$
in the $(t,x)$ plane occupied by the string, so that the finite quantity is the
transition probability per unit time (the rate) and per unit length of the
string.

The formula in Eq.(\ref{rz}) corresponds to a calculation of the decay rate as
the imaginary part of the energy of the initial string. At a finite temperature
$T$ the corresponding relevant quantity is the imaginary part of the free
energy\cite{Langer}, which one can calculate in the Euclidean space by
considering periodic in time configurations with the period $\beta = T^{-1}$. In
other words the thermal calculation corresponds to the path integration in the
Euclidean space-time having the topology of a cylinder. The nucleation rate is
then described by the same formula (\ref{rz}) with the action and the area being
calculated over one period, ${\cal A}=X \, \beta$.

\begin{figure}[ht]
  \begin{center}
    \leavevmode
    \epsfxsize=7cm
    \epsfbox{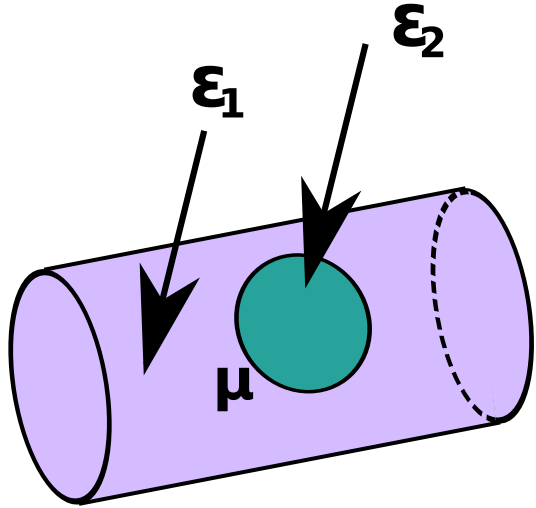}
    \caption{Bounce configuration for nonzero temperature}
  \end{center}
\label{bounce}
\end{figure}

In the present paper we consider only sufficiently small temperatures
$\b>\ell_c=2\,R$, at which temperatures we find the thermal effects behaving as
powers of $T$, which distinguish the string process from the decay of metastable
vacuum\cite{Garriga94}. We also treat the length $X$ of the metastable string as
the largest length parameter in the problem, so that $\b \ll X$. Under these
conditions the bounce corresponding to the action (\ref{a0}) is the same as at
zero temperature, except that it is placed on a cylinder rather than on a large
flat plane (Fig. \ref{bounce}).

We aim at calculating the path integral over the variations of the string around
the bounce configuration, which involves in particular the integration over the
shifts of the string in transverse directions, i.e. those perpendicular to $x$.
The contribution from each of the $d-2$ transverse dimensions factorizes, so
that it is sufficient to consider only one transverse dimension with further
straightforward generalization to an arbitrary dimension. The coordinates on the
cylinder (or on the plane where all the points separated by $n\b$ along
Euclidean time are identified) are $t$, $x$, with $t$ being the periodic time
coordinate, and the coordinate orthogonal to the surface of the cylinder is $z$.
The boundary conditions for the configurations over which we integrate are
\be
z \left ( t,x= \pm {X \over 2} \right )=0, \,~~~~~z(t+\b,x)=z(t,x)~.
\label{boundary_conditions}
\ee
Introducing the polar coordinates $(r,\theta)$ in the $(t,x)$-plane one can
consider small variations of the classical configuration (Figs. 2, 3) in the
following form. Variations of the interface are given by
\be
r(\theta) = R + f(\theta)\,~~~~z(R, \theta)=\xi(\theta)~,
\ee
while the variations of the surfaces of the string in the bulk are
$z_1(r,\theta)$ and
$z_2(r,\theta)$ with the boundary conditions on the interface
\be
z_1(R,\theta)= z_2(R,\theta)=\xi(\theta).
\label{boundary_zeta}
\ee
One can certainly notice that the symmetry of the polar coordinates is that of
the disk describing the bounce, but it is not the symmetry of the periodicity
condition in $t$. It will be clear from the subsequent calculation, that in fact
all the discussed thermal effect can be viewed as originating from this
mismatch, once it is properly accounted for.

\begin{figure}[ht]
  \begin{center}
    \leavevmode
    \epsfxsize=7cm
    \epsfbox{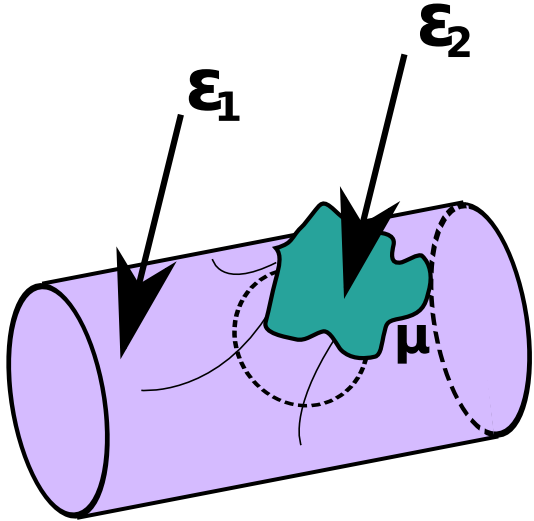}
    \caption{Variation of the bounce configuration}
  \end{center}
\label{variation}
\end{figure}
In terms of the introduced variables the action (\ref{a0}) can be written in the
quadratic approximation in the deviations from the bounce as
\bea
&&S_{12}=\eps_1  \, X \, \beta + {\pi \, \mu^2 \over \eps_1- \eps_2} +
{\eps_1-\eps_2 \over 2} \, \int d\theta \, \left (\dot{\xi}^2+\dot{f}^2-f^2
\right )+
\nonumber \\
&& {\eps_1 \over 2} \, \int_{\rm out} r drd\theta \,\left ( z_1'\,^2+
{\dot{z_1}^2
\over r^2} \right ) + {\eps_2 \over 2} \, \int_{\rm in} r drd\theta \,\left (
z_2'\,^2+ {\dot{z_2}^2 \over r^2} \right )~,
\label{s12}
\eea
where the primed and dotted symbols stand for the derivatives with respect to
$r$ and $\theta$ correspondingly, and the ``out" and ``in" integrals are taken
over the surface respectively outside and inside the disk $r < R$.

Similarly, the action around a flat initial string configuration in the
quadratic
approximation takes the form
\be
S_1 = \eps_1 \, X \, \b  +{\eps_1 \over 2} \, \int r drd\theta \,\left (
z'\,^2+ {\dot{z}^2 \over r^2} \right )~,
\label{s1}
\ee
with the integral running over the whole space-time cylinder and $z(r,\theta)$
parametrizing small deviations of the string in the
transverse direction.

\section{Integration}

The path integral ${\cal Z}_1$ runs  all transverse fluctuations $z(r,\theta)$
with the action given by Eq.(\ref{s1}). This integral contains no interface at
$r=R$. However, as explained in Ref.~\cite{mv08}, for the purpose of calculation
of the ratio of the partition functions in Eq.(\ref{rz}) it is helpful to
organize the integration for ${\cal Z}_1$ in the following way. First one fixes
the values of $z$ on the circle of radius $R$: $z(R, \theta)=\xi(\theta)$, and
integrates over the remaining bulk variables with fixed $\xi(\theta)$. The full
partition function ${\cal Z}_1$ is then found after path integration over the
boundary variations $\xi(\theta)$. As demonstrated in Ref.~\cite{mv08} this
procedure in fact factorizes the full path integral into a product of ``bulk"
and ``boundary" factors with the bulk factors being the same in ${\cal Z}_{12}$
and ${\cal Z}_1$. As a result the ratio of the full path integrals in
Eq.(\ref{rz}) is determined only by the boundary terms:
\be
{d \Gamma \over d \ell} = {\eps_1-\eps_2 \over 2 \pi} \, \exp \left ( -{\pi
\,\mu^2 \over \eps_1-\eps_2} \right ) \, \, {{\cal Z}_{12{\rm (boundary)}} \over
{\cal
Z}_{1{\rm (boundary)}}} ~
\ee
with the boundary partition functions ${\cal Z}_{12 (\rm boundary)}$ and ${\cal
Z}_{1 (\rm boundary)}$ being given by
\bea
{\cal Z}_{12 (\rm boundary)}=\int {\cal D}\xi \, \exp \left [ - {\eps_1-\eps_2
\over 2} \, \int d\theta \, \dot{\xi}^2 - R \, \int d \theta  \,  \left (
{\eps_2 \over 2} {\partial_r} z_{2c} \left . \right |_{r=R} - {\eps_1 \over 2}
{\partial_r} z_{1c} \left . \right |_{r=R} \right ) \, \xi \right]
\label{partition_z12} \nonumber \\
{\cal Z}_{1 (\rm boundary)}=\int {\cal D}\xi \, \exp \left [ -R \,{\eps_1
\over 2} \, \int d \theta \,  \left ({\partial_r} z_{2c} \left . \right |_{r=R}
-  {\partial_r} z_{1c} \left . \right |_{r=R} \right ) \, \xi \right]~,
\label{partition_functions}
\eea
where the functions $z_{1c}$ and $z_{2c}$ satisfy the Laplace equation $\Delta
z_{ic}=0$ with the boundary conditions
\be
z_{1c}(R,\t)=z_{2c}(R,\t)=\xi(\t)~.
\label{bc12}
\ee
The outer solution $z_{1c}$ is also periodic in time
$z_{1c}(t+\b,x)=z_{1c}(t,x)$
and satisfies the zero boundary condition at the spatial infinity
\be
z_{1c} \left (t,x=\pm {X \over 2} \right )=0
\label{bcz1}
\ee
while the inner solution $z_{2c}$ is required to be regular inside the disk.

In order to do the path integrals one can expand the boundary function
$\xi(\theta)$ in angular harmonics:
\be
\xi(\theta)={a_0 \over \sqrt{2\pi}} + {1 \over \sqrt{\pi}}\sum_{n=1}^\infty
\left [ a_n \, \cos( n \theta)+b_n \, \sin(n \theta) \right ].
\label{ft}
\ee
For the inner solution $z_{2c}$ to the Laplace equation, i.e. at $r \le R$ one
finds no difficulty in finding the harmonics matching the boundary function at
the interface\cite{mv08}:
\be
z_{2c}^{(0)}= {a_0 \over \sqrt{2\pi}}\,,~~~~z_{2c}^{(n)}(r,
\theta)= {1 \over \sqrt{\pi}} \, \left [ a_n \, \cos( n \theta)+b_n \, \sin(n
\theta) \right ] \, {r^n \over R^n}~.
\ee

For the outer solution $z_{1c}$ however there is a difficulty due to the
previously mentioned mismatch between the symmetry of the boundary and of the
periodicity conditions. It is impossible to choose the solution to the Laplace
equation for outer string bulk variable to be $z_{1c}^{(n)}(r,\t)\sim r^{-n}$,
since it is not periodic in time. In this situation in order to have a periodic
solution one can perform a periodic mapping of the cylinder on the plane and
consider the outer solution of the Laplace equation as the sum of the solutions
produced by a ``source" at each period as illustrated in Fig.~4. Introducing the
complex variable $w=t+i\,x$, we construct the solutions for the functions
$z_{1c}(r,\theta)$ using the harmonic real and imaginary parts of the following
basis set of periodic functions, satisfying the boundary condition (\ref{bcz1})
at large $|x|$,
\bea
g_0(w)&=&\ln\l[\sin\l(\f{\pi\,w}{\b}\r)\r] - { \pi \, X \over 2 \b} - i \, {\pi
x \over X} + \ln 2~, \nonumber \\
g_1(w)&=&\f{\pi\,R}{\b}\,\cot\l(\f{\pi\,w}{\b}\r) +i \, {2\, \pi \, R \,x \over
\b \,X}~,\nonumber \\
{g}_k(w)&=&\f{R^k}{w^k}+\sum_{n=1}^\infty\l[\f{R^k}{(w-n\b)^k}+\f{R^k}{(w+n\b)^k
}\r],~~~\textrm{for $k>1$}~.
\eea
\begin{figure}[ht]
  \begin{center}
    \leavevmode
    \epsfxsize=10cm
    \epsfbox{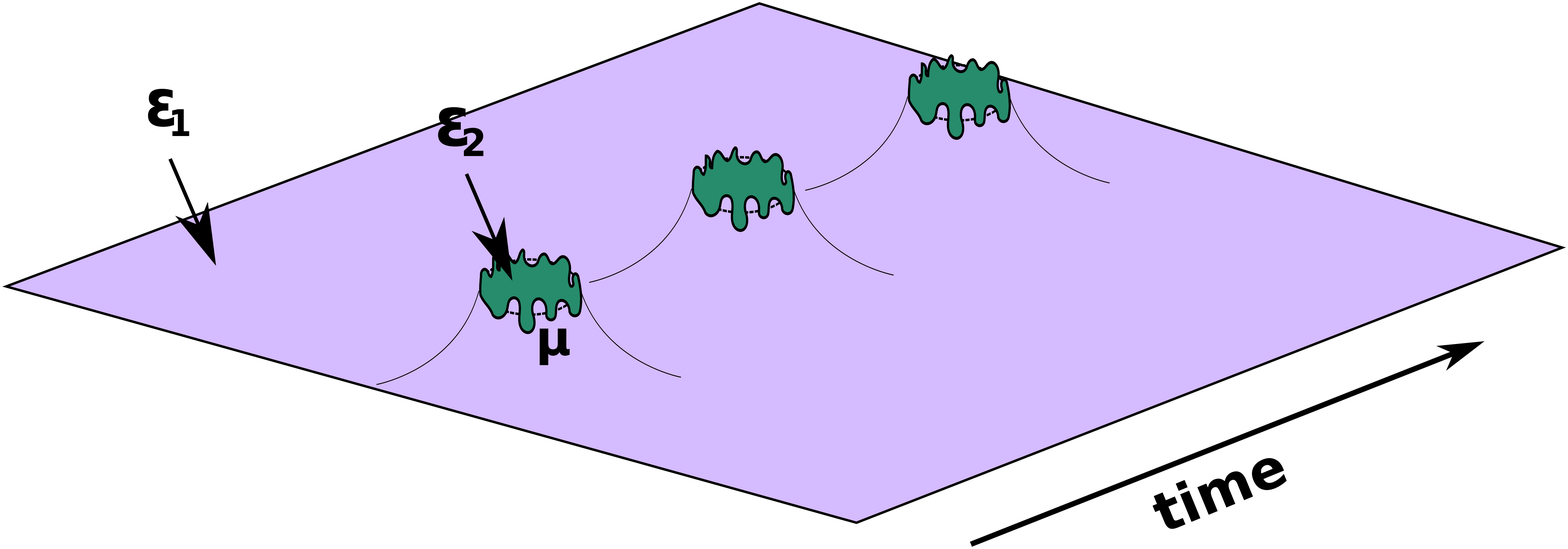}
    \epsfxsize=6cm
    \epsfbox{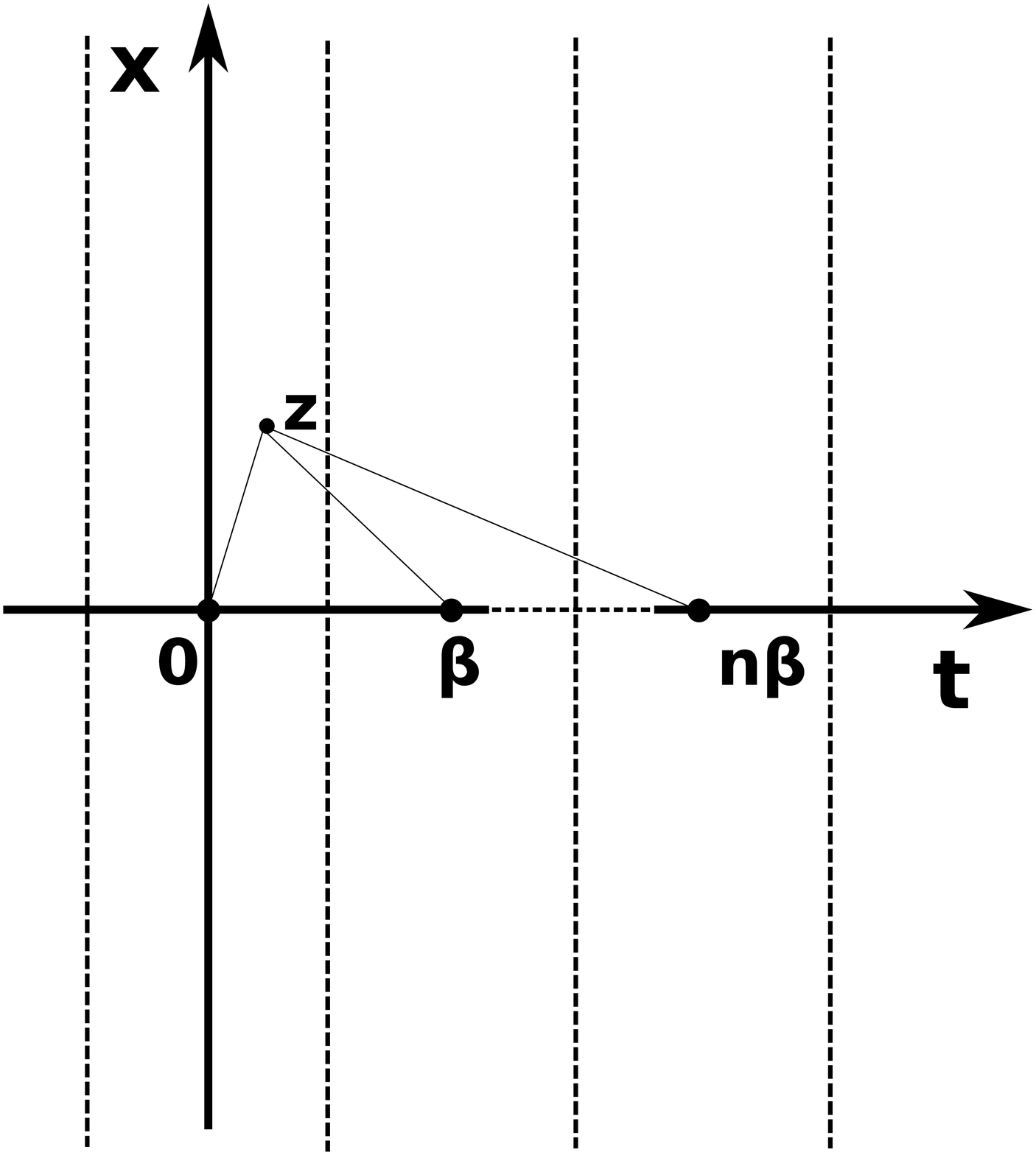}
    \caption{Periodic configuration on the plane}
  \end{center}
\label{periodic_plane}
\end{figure}

Clearly these functions are periodic in the Euclidean time by construction, with
the period $\b$. Also the functions $g_k(w)$ with $k > 1$ are analytic complex
functions, so that their real and imaginary parts are harmonic. In the functions
$g_0(w)$ and $g_1(w)$ the explicit dependence on $x$, introducing
non-analyticity, is linear and is thus also harmonic, so that their real and
imaginary parts do satisfy the Laplace equation.

An arbitrary periodic outer solution to the Laplace equation, satisfying the
boundary condition (\ref{bcz1}) at large $|x|$ can be expanded in the series
\be
z_{1c}(r,\t)=A_0\mathrm{Re}\,g_0(w)+\sum_{k=1}^\infty
A_k\mathrm{Re}\,g_k(w)+B_k\mathrm{Im}\,g_k(w),
\ee
The disadvantage of this set of solutions (hence of such an expansion) is that
it is not orthogonal, so that the expression for the action up to quadratic
terms is not diagonal in this basis. Therefore, the calculation of the integral
is not just a calculation of the product of eigenvalues. To find the integral
over the amplitudes of the Fourier harmonics one has to express the amplitudes
of the solutions chosen $A_k$,$B_k$ in terms of the amplitudes $a_k$, $b_k$. The
relation between the coefficients $A_k$,$B_k$ and $a_k$, $b_k$ can be found from
the matching condition (\ref{bc12}) on the interface
\be
{a_0 \over \sqrt{2\pi}} + {1 \over \sqrt{\pi}}\sum_{n=1}^\infty
\left [ a_n \, \cos( n \theta)+b_n \, \sin(n \theta) \right
]=A_0\,g_0(w)\Big|_{r=R}+\sum_{k=1}^\infty\l[A_k\mathrm{Re}g_k(w)+B_k\mathrm{Im}
g_k(w)\r]\Big|_{r=R}.
\ee
One can readily notice that the function $g_0$ contains a large constant term,
proportional to $X$, which totally dominates the matching condition for the
$a_0$ mode at $r=R$, so that
\be
{a_0 \over \sqrt{2\pi}}=-A_0\,\f{\pi\,X}{2 \b}\,\l[1+O(R/X)\r].
\label{a0A0}
\ee
For this reason the effect of the mixing between $a_0$ and higher modes is
suppressed by inverse powers of $X$ and can be ignored in the limit of a long
string. For this reason in considering the mixing of the modes in the following
calculation we keep only $k\neq0$. Furthermore the linear in $x$ terms in the
functions $g_0$ and $g_1$ are suppressed at $r=R$ by the factor $R/X$ and we
also neglect them.

In what follows we consider the expansion of the functions $g_k$ at $r=R$ in
powers of $R/\b$, which expansion, as will be seen later, converges at $R <
\b/2$.
Using
\be
(1+x)^{-k}=1+\sum_{p=1}^{\infty}(-1)^px^pC_{p+k-1}^p,
\ee
where $C_{p+k-1}^p$ are binomial coefficients,
\be
C_{p+k-1}^p=\f{(p+k-1)!}{p!(k-1)!}
\ee
and also a definition of the Riemann $\zeta$-function
\be
\zeta(k)=\sum_{n=1}^\infty n^{-k}~,
\ee
we find for $k\neq0$
\be
g_k(w)=\f{R^k}{w^k}+\sum_{p=1}^\infty d_{pk}\l(\f{w}{R}\r)^p,
\label{g_exp}
\ee
with
\be
d_{pk}=\l[(-1)^k+(-1)^p\r]\l(\f{R}{\b}\r)^{k+p}\,\zeta(p+k)\,C_{p+k-1}^p.
\label{D_matrix}
\ee
We have omitted in the expression (\ref{g_exp}) a constant term, which describes
the mixing with the $a_0$ mode as well as a term explicitly proportional to
$R/X$.
Using the expansion (\ref{g_exp}) for $g_k(w)$ and considering the real part of
the functions $g_k(w)$, one can find the coefficients $a_l$ in terms of $A_k$
\be
a_l=\f{1}{\pi}\sum_kA_k\int_0^{2\pi}\mathrm{Re}g_k\cos(l\t)d\t=
A_l+\sum_{k=1}^\infty A_k\,d_{lk}~,
\ee
or in the matrix form
\be
A=(1+D)^{-1}\, a~,
\label{cos_solution}
\ee
where matrix $D$ has elements $d_{lk}$. Similarly for the imaginary part one
gets
\be
b_l=\f{1}{\pi}\sum_kB_k\int_0^{2\pi}\mathrm{Im}g_k\cos(l\t)d\t=B_l-\sum_kB_kd_{l
k}
\ee
and
\be
B=(1-D)^{-1} \,b~.
\ee
As usual, the contributions from $\cos$ and $\sin$ modes are independent, and we
consider the contribution to the boundary term in (\ref{partition_z12}) from the
even ($\cos$) harmonics first
\bea
&&-\int_0^{2\pi}\l.\l(Rg^{(R)}\p_rg^{(R)}\r)\r|_Rd\t=\int_0^{2\pi}d\t\l[\sum_la_l\
cos(l\t)\r]
\sum_k\l[k\l(\f{R}{r}\r)^k\cos(k\t)-\sum_{p=1}^\infty pd_{pk}\r] = \nonumber \\
&& \sum_kA_k\l[ka_k-\sum_ppa_pd_{pk}\r]=
\sum_{k,p}a_p\l[k\delta_{pk}-pd_{pk}\r]A_k~.
\label{boundary_term_sum}
\eea
Introducing the matrix
\be
\hat{N}=\mathrm{diag}(1,2,\dots,n,\dots)~,
\ee
one can rewrite the expression (\ref{boundary_term_sum}) in the matrix form
\be
-\int_0^{2\pi}\l.\l(Rg^{(R)}\p_rg^{(R)}\r)\r|_Rd\t=a\hat{N}(1-D)A~.
\ee
A substitution in this expression of the solution for $A$ in terms of $a$
(\ref{cos_solution}) leads to
\be
-\int_0^{2\pi}\l.\l(Rg^{(R)}\p_rg^{(R)}\r)\r|_Rd\t=a\hat{N}(1-D)(1+D)^{-1}a~.
\ee
Clearly, for the odd ($\sin$) modes one gets the same expression with the
replacement $D\to-D$. Collecting all the terms together one can write the result
for the boundary partition functions (\ref{partition_functions}) as
\bea
{\cal Z}_{12 (\rm
boundary)}&=&\mathrm{Det}\l[(\e_1-\e_2)\hat{N}^2+\e_2\hat{N}+
\e_1\hat{N}(1-D)\f{1}{1+D}\r]^{-1/2}\cdot\{D\to-D\} \nonumber \\
{\cal Z}_{1 (\rm
boundary)}&=&\mathrm{Det}\l[\e_1\hat{N}+\e_1\hat{N}(1-D)\f{1}{1+D}\r]^{-1/2}
\cdot\{D\to-D\},
\label{z12z1d}
\eea
where $\{D\to-D\}$ means that one should take the expression and make the
replacement $D\to-D$.

The zero temperature limit for the probability rate $d\Gamma_0/d\ell$ formally
corresponds to setting $D \to 0$. Thus one can use the known result for the zero
temperature decay rate \cite{mv08}, and concentrate on a calculation of the
thermal catalysis factor $K$ defined as
\be
\f{d\Gamma}{d\ell}=K \, \f{d\Gamma_0}{d\ell}\,.
\ee
In a $d$-dimensional theory, i.e. with $d-2$ transverse dimensions, the catalysis factor can be written as $K=G^{d-2}$, where $G$ is the factor per each transverse direction given by
\be
G=
\f{{\cal Z}_{12 (\rm
boundary)}}{{\cal Z}_{12 (\rm boundary)}^{D=0}}
\,
\f{{\cal Z}_{1 (\rm boundary)}^{D=0}}{{\cal Z}_{1 (\rm boundary)}}\,.
\ee
According to Eq.(\ref{z12z1d}) it is a matter of simple algebra to express the
factor $G$ in terms of the matrix $D$:
\be
G=\Det\l[1-\l(\f{\hat{N}-1}{\hat{N}+b}\,D\r)^2\r]^{-1/2}
\label{gen_formula}
\ee
where we have introduced the parameter $b=\dst\f{\e_1+\e_2}{\e_1-\e_2}$.

\section{Analysis of the general formula}
In this section we consider in some detail the temperature effect in the string
transition rate described by our general formula (\ref{gen_formula}) in the
situation where the inverse temperature is larger then the diameter of the
classical configuration (bounce) $\b>2R$. We first notice that due to the
presence of the factor $(\hat{N}-1)$ the first elements from the first row and
the first column of the matrix $D$, $d_{1k}$ and $d_{p1}$, enter the  expression
$[(\hat{N}-1) \, D]^2$ with zero coefficients, so that the final result
(\ref{gen_formula}) does not depend on them. Furthermore, one can see from
Eq.(\ref{D_matrix}) that the matrix element $d_{pk}$ is not equal to zero only
if the indices $p$ and $k$ have the same parity. Hence, there is no mixing
between the amplitudes with even ($a_{2l}$, $b_{2l}$) and odd ($a_{2l+1}$,
$b_{2l+1}$) indices. Therefore the determinant in the (\ref{gen_formula}) can be
written as a product of determinants corresponding to even and odd amplitudes
\be
\Det\l[1-\l(\f{\hat{N}-1}{ \hat{N}+b}\,D\r)^2\r]=\Det\l[1-\l(\f{2 \, 
\hat{N}-1}{\hat{2 \, N}+b}\,U\r)^2\r]\,
\Det\l[1-\l( \f{2 \, \hat{N}}{2 \, \hat{N}+1+b}\,V\r)^2\r],
\label{uvd}
\ee
where the matrix elements of the matrices $U$ and $V$ are
\be
U_{pk}=  d_{2p\,2k} \,,~~~~~
V_{pk}=d_{2p+1,\,2k+1}~,
\ee
and the indices $p$ and $k$ take values $1,2,\ldots$\,.

For practical calculations it is also convenient to write the expressions for
the elements of the matrices entering in Eq.(\ref{uvd}) in terms of their
indices:
\bea
\l(\f{2 \,  \hat{N}-1}{\hat{2 \, N}+b}\,U\r)_{pk} &=& 2 \, {2 p - 1 \over 2 p
+b} \, {(2p+2 k -1)! \over (2p)! \, (2 k-1)!} \, \zeta(2p+2k) \, \left ( {R
\over \beta} \right )^{2p+2k}~, \nonumber \\
-\l( \f{2 \, \hat{N}}{2 \, \hat{N}+1+b}\,V\r)_{pk} &=& {4 p \over 2p+1 +b} \,
{(2p+2k+1)! \over (2p+1)! \, (2 k)!} \, \zeta(2p+2k+2) \, \left ( {R \over
\beta} \right )^{2p+2k+2}~,
\label{uvme}
\eea
with $p,k=1,2,\ldots$.

In order to find the first thermal correction at low temperature one can  expand
the matrices in power series using well known formula for the determinant
\be
\mathrm{Det}(1-M)=\exp\l[\mathrm{Tr}\ln\l(1-M\r)\r]=\exp\l[-\mathrm{Tr}\sum_{l=1
}^\infty\f{M^l}{l}\r]=1
-\mathrm{Tr}M+O(M^2).
\ee
In our case
\be
\l(\f{2 \, \hat{N}-1}{2 \, \hat{N}+b}\,U\r)^2=\l(
\ba{ccc}
\dst
\f{36\,\zeta^2(4)}{(2+b)^2}\l(\f{R}{\b}\r)^8+\f{600\,\zeta(6)^2}{(2+b)\,(4+b)}\,
\l(\f{R}{\b}\r)^{12} & \dst
\f{120\,\zeta(4)\zeta(6)}{(2+b)^2}\,\l(\f{R}{\b}\r)^{10} & \cdots \\ \dst
\f{180\,\zeta(4)\zeta(6)}{(2+b)\,(4+b)}\,\l(\f{R}{\b}\r)^{10} &
\dst\f{600\,\zeta(6)^2}{(2+b)\,(4+b)}\,\l(\f{R}{\b}\r)^{12} & \cdots \\
\vdots & \vdots & \ddots
\ea
\r)~,
\ee
\be
\l(\f{2 \, \hat{N}}{2\, \hat{N}+1+b}\,V\r)^2=\l(
\ba{cc}
\dst \dst\f{1600\,\zeta(6)^2}{(3+b)^2}\,\l(\f{R}{\b}\r)^{12} & \cdots \\
\vdots & \ddots
\ea
\r)~.
\ee
Therefore the first correction to the zero temperature value of the rate is
proportional to $R^8/\b^8$ and is given by Eq.(\ref{kf}).

\begin{figure}[ht]
  \begin{center}
    \leavevmode
    \epsfxsize=7cm
    \epsfbox{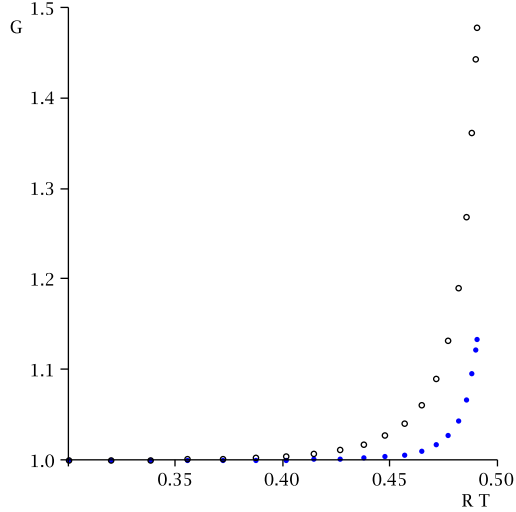}
    \caption{The thermal catalysis factor per each transverse dimension vs $RT$
for $b=1$ (circles) and for $b=10$ (solid circles)}
  \end{center}
\label{corrections}
\end{figure}
The series for the function $G(T)$ diverges when $\b=2\,R$. The corrections for
the temperature close to $(2R)^{-1}$ can be found numerically. The proximity to
this point defines the number of terms which should be taken into account in the
series for $G(T)$. The plot for the function $G(T)$ vs the parameter $R/\b=R\,T$
calculated numerically with 50 first rows and columns retained in the matrices
$U$ and $V$ is shown in Fig.~5.

\section{Summary}
In the present paper we considered the  transition between two states of a
string at finite temperature. The result for the rate of the transition in
arbitrary number of dimensions is obtained in a form of a product of the value
at zero temperature and the catalysis factor (\ref{gen_formula}). The numerical
analysis of the expression shows (see Fig.\ref{corrections}) that for smaller
$b$ the influence of the temperature on the process becomes noticeable earlier
and the catalysis factor grows more rapidly with temperature. Our result is
applicable in the range of temperatures for which the associated period in
Euclidean time is larger then the critical size of the classical configuration
$\ell_c\,T>1$, so that it is still possible to inscribe the bounce in one
period. We have shown that, similarly to the process of false vacuum decay at
finite temperature in two dimensions \cite{Garriga94}, there is no change in the
exponential behavior of the probability rate while the temperature is small
enough, and that such change starts only when $\b$ becomes smaller than the
diameter of the bounce. On the other hand we have demonstrated that unlike in
the case of false vacuum decay, there arise thermal corrections to the
transition rate at low temperature, which corrections have power  dependence on
the temperature and are due to thermal waves excited on the string at
arbitrarily low temperatures. In practical terms, as can be seen from Fig.~5,
the value of the catalysis factor changes from $1$ for $R\,T=0$ to approximately
$1.5$ for $R\,T=0.49$, i.e. quite slowly, even for $b=1$.  However there is an
actual singularity in this factor at $R\,T=1/2$ so that the string transition
becomes significantly accelerated by thermal effects for temperatures close to
$1/\ell_c$.

\section*{Acknowledgments}
This work is supported in part by  the DOE grant DE-FG02-94ER40823.



\begin{thebibliography}{99}
\bibitem{Vilenkin}
  A.~Vilenkin,
  Nucl.\ Phys.\  B {\bf 196}, 240 (1982).
\bibitem{Preskill}
  J.~Preskill and A.~Vilenkin,
  Phys.\ Rev.\  D {\bf 47}, 2324 (1993)
  [arXiv:hep-ph/9209210].
\bibitem{Shifman}
  M.~Shifman and A.~Yung,
  Phys.\ Rev.\  D {\bf 66}, 045012 (2002)
  [arXiv:hep-th/0205025].
\bibitem{vko}
  M.~B.~Voloshin, I.~Y.~Kobzarev and L.~B.~Okun,
  Sov.\ J.\ Nucl.\ Phys.\  {\bf 20}, 644 (1975)
  [Yad.\ Fiz.\  {\bf 20}, 1229 (1974)].
\bibitem{Coleman}
  S.~R.~Coleman,
  Phys.\ Rev.\  D {\bf 15}, 2929 (1977)
  [Erratum-ibid.\  D {\bf 16}, 1248 (1977)].

\bibitem{Schwinger}
  J.~Schwinger, Phys.\ Rev.\ {\bf 82}, 664 (1951).
\bibitem{Stone}
  M.~Stone,
  Phys.\ Rev.\  D {\bf 14}, 3568 (1976).

\bibitem{Callan}
  C.~G.~Callan and S.~R.~Coleman,
  Phys.\ Rev.\  D {\bf 16}, 1762 (1977).

\bibitem{Langer}
J.~S.~Langer, Ann.\ Phys.\ (N.Y.) {\bf 41}, 108 (1967).

\bibitem{mv08}
  A.~Monin, M.~B.~Voloshin,
  arXiv:0808.1693	
\bibitem{Garriga94}
  J.~Garriga,
  Phys.\ Rev.\ D {\bf 49}, 5497 (1994)





\end{thebibliography}
\end{document}